\documentclass[superscriptaddress, reprint, twocolumn%
reprint,
amsmath,amssymb,
aps,
prb,
]{revtex4-2}

\usepackage{standalone}

\begin{document}


	\preprint{APS/123-QED}
	
	\title{Imaging ultrafast electronic domain fluctuations with X-ray speckle visibility}
	
	\author{N. Hua}
	\email{nelson.hua@psi.ch}
	\affiliation{PSI Center for Photon Science, Paul Scherrer Institute, 5232 Villigen PSI, Switzerland}%
	\author{Y. Sun}
	\affiliation{Linac Coherent Light Source, SLAC National Accelerator Laboratory, Menlo Park, California 94025, USA}
	\author{P. Rao}
	\author{N. Zhou Hagstr\"{o}m}
	\affiliation{Department of Materials Science Engineering, University of California Davis, Davis, California 95616, USA}
	\author{B. K. Stoychev}
	\author{E. S. Lamb}
	\affiliation{Department of Physics, University of California San Diego, La Jolla, California 92093, USA}
	\author{M.~Madhavi}
	\author{S.~T. Botu}
	\affiliation{Department of Materials Science Engineering, University of California Davis, Davis, California 95616, USA}
	\author{S.~Jeppson}
	\affiliation{Department of Materials Science Engineering, University of California Davis, Davis, California 95616, USA}
	\author{M. Clémence}
	\author{A. G. McConnell}
	\affiliation{PSI Center for Photon Science, Paul Scherrer Institute, 5232 Villigen PSI, Switzerland}%
	\affiliation{Laboratory for Solid State Physics and Quantum Center, ETH Zürich, 8093 Zürich, Switzerland}
	\author{S.-W. Huang}
	\author{S.~Zerdane}
	\author{R. Mankowsky}
	\author{H. T. Lemke}
	\author{M. Sander}
	\affiliation{PSI Center for Photon Science, Paul Scherrer Institute, 5232 Villigen PSI, Switzerland}
	\author{V. Esposito}
	\author{P. Kramer}
	\author{D. Zhu}
	\author{T.~Sato}
	\author{S. Song}
	\affiliation{Linac Coherent Light Source, SLAC National Accelerator Laboratory, Menlo Park, California 94025, USA}
	\author{E. E. Fullerton}
	\affiliation{Center for Memory and Recording Research, University of California San Diego, La Jolla, California 92093, USA}
	\author{O. G. Shpyrko}
	\affiliation{Department of Physics, University of California San Diego, La Jolla, California 92093, USA}
	\author{R. Kukreja}
	\email{rkukreja@ucdavis.edu}
	\affiliation{Department of Materials Science Engineering, University of California Davis, Davis, California 95616, USA}
	\author{S. Gerber}
	\email{simon.gerber@psi.ch}
	\affiliation{PSI Center for Photon Science, Paul Scherrer Institute, 5232 Villigen PSI, Switzerland}%
	\date{\today}
	
	\begin{abstract}
		
\bf{Speckle patterns manifesting from the interaction of coherent X-rays with matter offer a glimpse into the dynamics of nanoscale domains that underpin many emergent phenomena in quantum materials. While the dynamics of the average structure can be followed with time-resolved X-ray diffraction, the ultrafast evolution of local structures in nonequilibrium conditions have thus far eluded detection due to experimental limitations, such as insufficient X-ray coherent flux. Here we demonstrate a nonequilibrium speckle visibility experiment using a split-and-delay setup at an X-ray free-electron laser. Photoinduced electronic domain fluctuations of the magnetic model material Fe$_3$O$_4$ reveal changes of the trimeron network configuration due to charge dynamics that exhibit liquid-like fluctuations, analogous to a supercooled liquid phase. This suggests that ultrafast dynamics of electronic heterogeneities under optical stimuli are fundamentally different from thermally-driven ones.}
	\end{abstract}
	
	\maketitle

Performing time-resolved X-ray scattering and spectroscopy experiments with femtosecond temporal resolution became feasible only recently with the advent of the X-ray free-electron laser (FEL). A plethora of nonequilibrium phenomena in quantum materials have since been demonstrated such as coherent control of lattice dynamics \cite{Mankowsky2014,Rettig2015,Gerber2015, Gerber2017, Lee2022}, enhancement of charge-density wave~(CDW) order \cite{Singer2016, Trigo2021, Wandel2022}, discovery of metastable phases~\cite{Lantz2017, Singer2018}, decoupling of structural and electronic order parameters~\mbox{\cite{Beaud2014, Rettig2019, Cammarata2021}}, and manipulation of spin dynamics in magnetic systems~\cite{Johnson2012,Graves2013, Zhou2022, Ueda2023, Jangid2023}. These experiments allow us to study not only collective phenomena in nonequilibrium conditions but open the possibility of using light to control material properties at ultrafast timescales \cite{delaTorre2021}. However, a critical drawback is that these techniques are blind to the dynamics of heterogeneities where fluctuations of an order parameter play a critical role in either suppressing or driving nonequilibrium phase transitions \cite{Johnson2023,Ko2023}, such as evolution of domain walls, vacancy and interstitial diffusion, as well as mobility of topological defects. For example, a dense network of domain walls induced by optical excitation is attributed to the metallic nature of the `hidden' CDW phase in \mbox{1\emph{T}-TaS$_{2}$}~\cite{Stojchevska2014}. Heterogeneities can also be intentionally introduced such as dopants in silicon, leading to highly entangled states for quantum computing paradigms that can by dynamically controlled in out-of-equilibrium conditions \cite{Crane2021}. In high-\emph{T$_{c}$} cuprates, it has been shown that inhomogenous charge- and spin-density wave stripe domains are intricately linked to the onset of superconductivity in La$_{2-x}$Sr$_{x}$CuO$_{4}$ \cite{Wen2019} while fluctuations of the charge order parameter seem to enhance superconductivity in La$_{2-x}$Ba$_{x}$CuO$_{4}$~\cite{Mitrano2019}. Therefore, understanding the interaction and evolution of heterogeneities in nonequilibrium conditions is a crucial step towards fine tuning photo-driven phenomena of correlated electron systems. 
	
	\begin{figure}
		\includegraphics[width=\columnwidth]{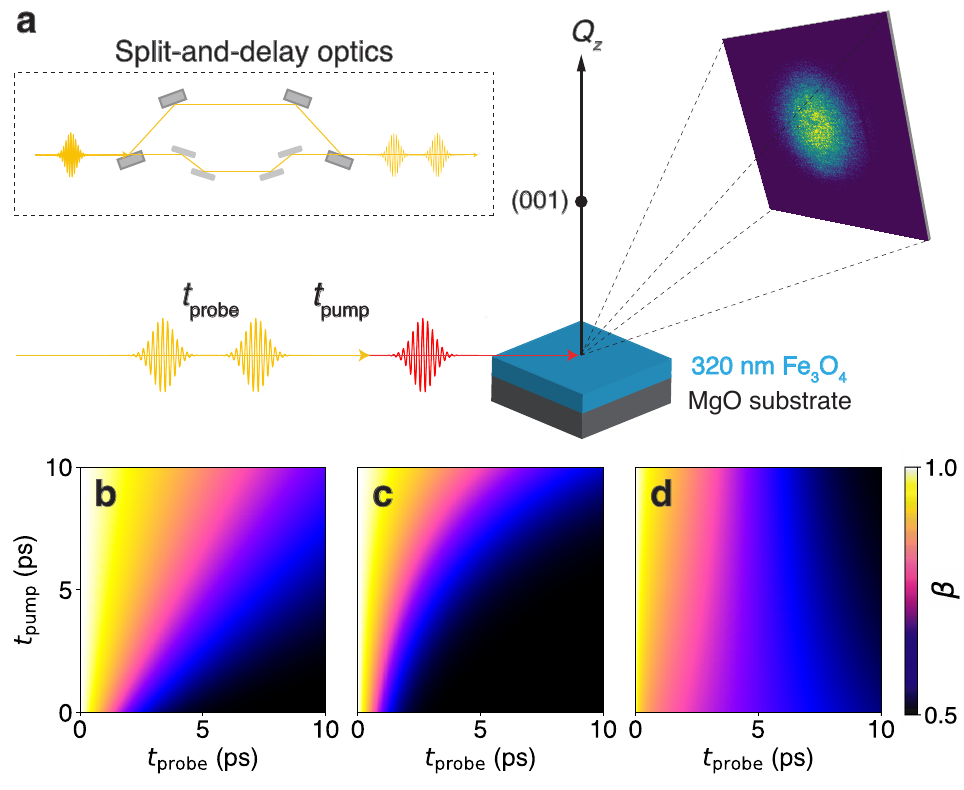}
		\caption{\label{Fig1} {\bf Pump-double-probe XSVS experimental setup.} {\bf a} Two coherent X-ray pulses delayed in time are produced from an SDO branch. An 800-nm optical excitation with a fluence of 1.5 mJ/cm$^{2}$ excites the Fe$_{3}$O$_{4}$ film followed by the two resonant (Fe $K$-edge) X-ray pulses that couple to the (001) charge order peak. Simulated speckle contrast maps for pump-double-probe XSVS show possible relaxation routes following optical excitation, such as {\bf b}~linear ($\tau$ $\propto$ \emph{t$_{\mathrm{pump}}$}) and {\bf c}~exponential ($\tau$ $\propto$ [exp(\emph{t$_{\mathrm{pump}}$})]) decay times. The type of dynamics can also evolve as shown in~{\bf d} where the system undergoes a glassy ($\alpha < $ 1) to a jammed~($\alpha >$ 1) transition after a critical \emph{t$_{\mathrm{pump}}$} value.}
	\end{figure}

	Techniques such as atomic and magnetic force, as well as scanning tunneling microscopy provide direct images of nanoscale structural and electronic heterogeneities. But they are surface sensitive and, therefore, lack information about the bulk. Recent technological advancements have extended these microscopy techniques down to picosecond temporal resolution \cite{Mogi2024, Liang2023}, but the scanning nature limits accessibility to reversible dynamics in a stroboscopic pump-probe setup. With coherent X-rays, it is possible to bypass these obstacles by probing the bulk while coupling to the dynamics of an order parameter that may be irreversible. For example, X-ray photon correlation spectroscopy (XPCS) has been used to directly measure spin-density wave domain fluctuations in chromium~\cite{Shpyrko2007}, inhomogenous interactions of CDW stripes in nickelates~\cite{Campi2022}, and ferroelectric nanodomain fluctuations in oxide heterostructures \cite{Zhang2017, Li2020}. XPCS extracts dynamical information by comparing changes in successive speckle patterns on the detector through an intensity-intensity autocorrelation analysis that is well suited for slow dynamics in equilibrium conditions. However, this type of analysis breaks down for dynamics that are faster than the detector frame rate, limited to the MHz range \cite{Allahgholi2019}, precluding access to sub-nanosecond regimes where fundamental light-matter interactions reside. To overcome this limitation, a modified coherent X-ray technique known as X-ray speckle visibility spectroscopy~(XSVS) can be used where the decay of the speckle pattern contrast on a single detector image is linked to the dynamic timescales in the system. This can be achieved by varying the X-ray pulse duration, as demonstrated for investigating caging effects of water \cite{Perakis2018}, but the temporal window is limited to tens to hundreds of femtoseconds. A less restrictive method employs split-and-delay optics (SDO) to produce two coherent X-ray pulses delayed in time (see inset of Fig.~\ref{Fig1}a) to probe dynamics on nano- to femtosecond timescales. In the latter case, it is assumed that the system is effectively static within the X-ray pulse duration \cite{Hruszkewycz2012} but dynamic between the two pulses. While the use of an SDO to perform XSVS experiments has been envisioned well before the first X-ray FEL \cite{Grubel2007, Gutt2007}, various inherent experimental difficulties have stunted its potential to become widely implemented. 
	
	An ideal XSVS experiment with a SDO produces two coherent X-ray pulses with equal intensities that propagate colinearly to ensure the same sample volume is probed. Any deviations lead to an artificial reduction of the initial speckle contrast that is needed to quantify the dynamic timescale. Several variations of SDO branches have been developed over the years~\mbox{\cite{Roseker09, Roseker2011, Osaka16, Zhu2017, Lu2018}}, but inherent experimental imperfections, specifically pointing instabilities and an angular mismatch between the two pulses, can be mistaken for dynamics or lead to a loss of the initial speckle contrast that jeopardizes the fidelity of the measurement. Analytical methods have been proposed \cite{Hua2020, Sun2020} to partially correct for these experimental flaws, but have thus far not been implemented successfully. Moreover, ultrafast dynamics in equilibrium conditions tend to exist in liquid systems~\cite{Roseker2018, Shinohara2020} with weak scattering signals that further exacerbate the photon counting statistics. Structural and electronic orders in solid-state systems, on the other hand, manifest as sharp diffraction peaks several magnitudes more intense than diffuse, ring-like scattering signatures from liquid systems, but generally exhibit slow dynamics in equilibrium conditions that can be studied with XPCS~\cite{Shpyrko2007, Campi2022, Zhang2017, Su2012, Sinha2014}. However, when these systems are excited to out-of-equilibrium conditions, dynamics of heterogeneities and defects enter the sub-nanosecond regime~\cite{Mitrano2019, deJong2013, Jangid2023,Johnson2023} that can be probed in the time domain via XSVS. 
 
 Using the spatial autocorrelation method proposed in~\cite{Sun2020} and by incorporating an optical pump with SDO, we developed a pump-double-probe experimental scheme (tr-XSVS \footnote{XSVS is inherently time-resolved. tr-XSVS refers to a nonequilibrium experiment where the time-resolution exists between the pump and the first coherent X-ray pulse}) to directly image ultrafast fluctuations of an order parameter in a solid-state system. Specifically, we investigate fluctuations of trimerons \cite{Senn2012}, the electronic order parameter defining the low-temperature insulating state of the model magnet Fe$_{3}$O$_{4}$, in a nonequilibrium insulator-metal (Verwey) transition triggered by optical excitation. Fe$_{3}$O$_{4}$ was one of the earliest materials where not only a first-order metal-insulator transition was experimentally confirmed, but also a charge-ordered phase was proposed to describe the insulating state \cite{Verwey1939}. Therefore, Fe$_{3}$O$_{4}$ is an archetypal material to study ultrafast nanoscale domain dynamics that can be extended to a wide class of materials exhibiting complex ordering and emergent behaviors.
 
\begin{figure*}
	\includegraphics[width=1\textwidth]{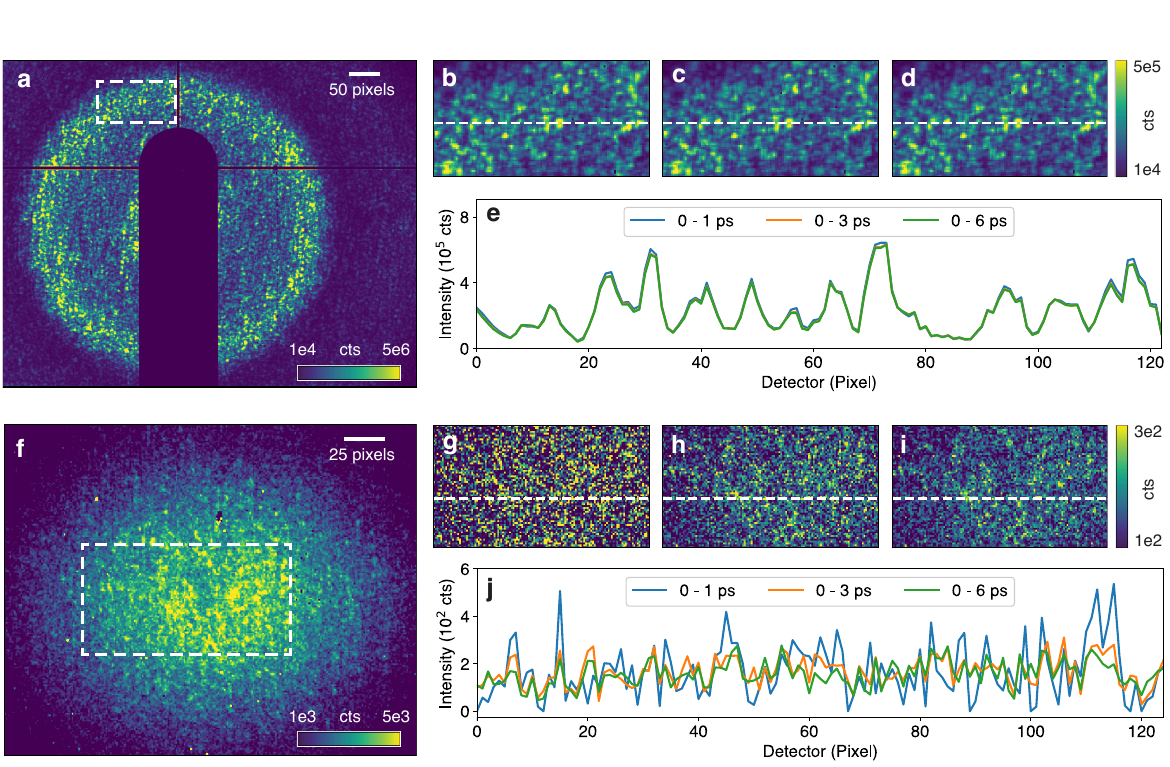}
	\caption{\label{Fig2} {\bf SDO stability and evidence for ultrafast charge dynamics.} Integrated detector image of a SiO$_{2}$ powder calibrant ($\approx$ 40,000 shots, {\bf a}) and the Fe$_{3}$O$_{4}$ (001) charge order peak ($\approx$ 70,000 shots, {\bf f}) measured with a fixed \emph{t$_{\mathrm{probe}}$} = 1 ps and a variable \mbox{$\emph{t$_{\mathrm{pump}}$} = 0 - 10$~ps}. The integrated region of interest defined by the white dotted boxes in {\bf a} and {\bf f}, binned by \emph{t$_{\mathrm{pump}}$} delays, is shown in {\bf b-d} and {\bf g-i} for SiO$_{2}$ and  Fe$_{3}$O$_{4}$, respectively. The static speckle contrast of SiO$_{2}$ independent of binned \emph{t$_{\mathrm{pump}}$} ranges in {\bf b} ($0 - 1$~ps), {\bf c} ($0 - 3$~ps),  and {\bf d}~($0 - 6$~ps) indicates the absence of dynamics. On the other hand, the decay of contrast, i.e. the variance of the intensity signal, of the Fe$_{3}$O$_{4}$ (001) charge order peak is clearly seen in {\bf g} ($0 - 1$~ps), {\bf h} ($0 - 3$~ps), and {\bf i}~($0 - 6$~ps), providing evidence for photoinduced domain changes. {\bf e} and {\bf j} show horizontal line cuts (white dashed lines) through the speckle patterns in {\bf b-d} and  {\bf g-i}, respectively. The vertical, dark feature in \textbf{a} is the shadow from a beamstop, blocking the high-intensity direct beam.}
\end{figure*}

\begin{figure*}[!ht]
	\includegraphics[width=1\textwidth]{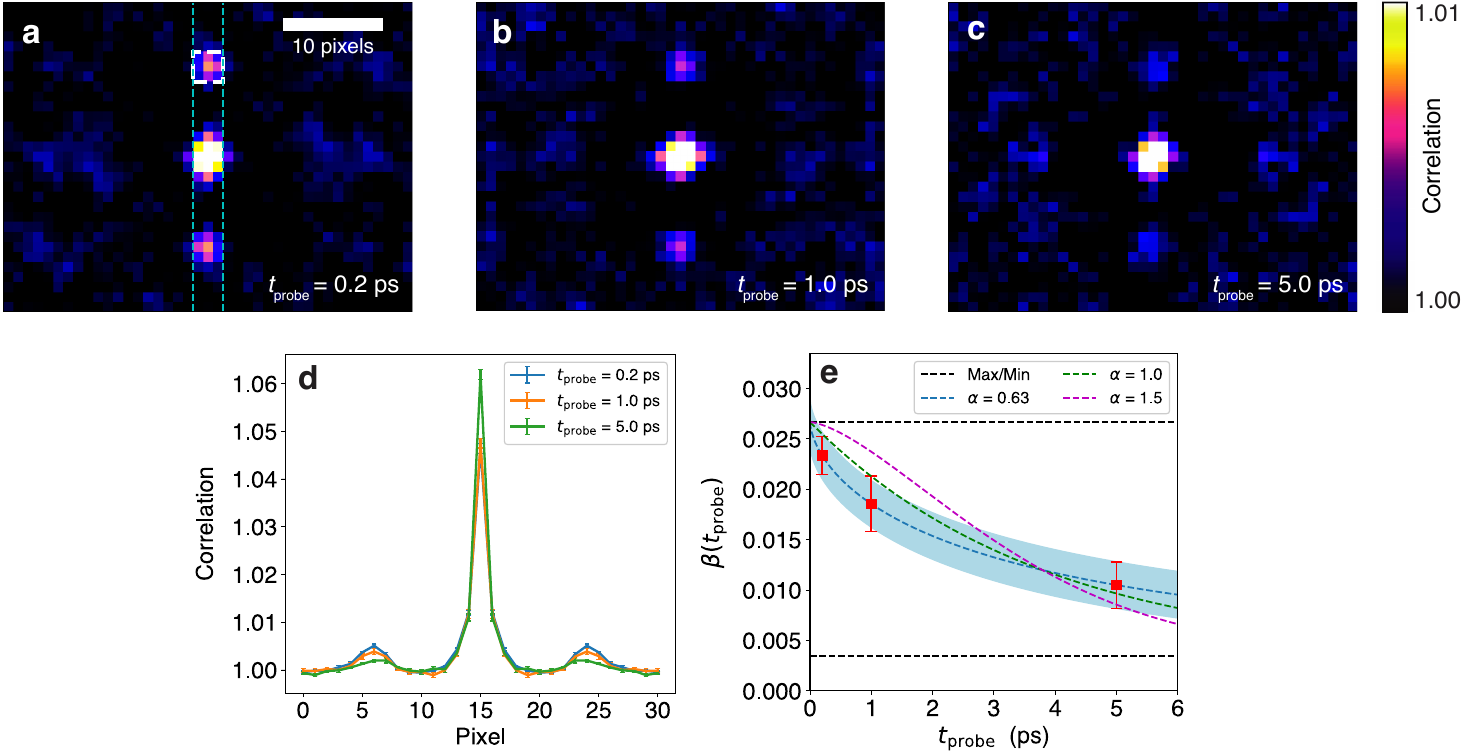}
	\caption{\label{Fig3} {\bf Spatial autocorrelation analysis and relaxation  timescale of fluctuations.} The spatial autocorrelation signal of the Fe$_{3}$O$_{4}$ (001) speckle pattern averaged over all \emph{t$_{\mathrm{pump}}$} delays for fixed \emph{t$_{\mathrm{probe}}$} delays of  0.2~({\bf a}), 1.0~({\bf b}), and 5.0~ps~({\bf c}). The autocorrelation side lobe is indicated by the white dotted box in {\bf a}. {\bf d} Vertical 1D projections of the area between the dotted cyan lines of {\bf a-c}. {\bf e} $\beta({t_\mathrm{probe}})$ extracted from the average side lobe intensities in~{\bf a-c} reveals a decay of speckle contrast. A model based on a stretched exponential decay of the intermediate scattering function points to glassy dynamics where the shaded blue region is centered on a decay defined by $\alpha = 0.63$ and $\tau = 3.79$~ps. References for Brownian ($\alpha$ = 1.0) and jammed ($\alpha$ = 1.5) dynamics with the same relaxation timescale are shown in green and purple, respectively.}
\end{figure*}
 
\section*{Results}
\subsection*{Pump-double-probe setup}
	
In an XSVS experiment with two coherent X-ray probes, the speckle contrast \emph{$\beta(t_{\mathrm{probe}})$} is directly related to the intermediate scattering function \emph{S}($Q,t_{\mathrm{probe}}$) by
\begin{equation}
	\label{Eq1}
	\beta(t_{\mathrm{probe}}) = \frac{\beta_{0}}{2}(1 + |{S(Q,t_{\mathrm{probe})}|^{2})},
\end{equation}
 where \emph{t}$_{\mathrm{probe}}$ is the time delay between the two X-ray pulses and \emph{$\beta_{0}$} is the initial contrast. Inherent dynamics of the system can be traced in the intermediate scattering function that follows $|{S(Q,t_{\mathrm{probe}})}| = \exp[-(t_{\mathrm{probe}}/\tau)^{\alpha}]$, where $\tau$ is the characteristic fluctuation timescale and the exponent $\alpha$ defines whether the nature of dynamics is glassy ($\alpha <1$), Brownian ($\alpha = 1$), or jammed ($\alpha > 1$). Here the time-resolved speckle contrast on the detector is the observable that gives access to the system's dynamics.  
		
For pump-double-probe tr-XSVS (see Fig.~\ref{Fig1}a) there exists an additional time delay between the optical pump and the first X-ray probe defined as \emph{t$_{\mathrm{pump}}$}. Speckle contrast maps can therefore be constructed from~\emph{$\beta(t_{\mathrm{pump}}, t_{\mathrm{probe}})$}, as shown in Fig.~\ref{Fig1}b-d, that resemble two-time correlation plots in XPCS to fully describe the nature of domain dynamics in nonequilibrium conditions. In the \emph{t$_{\mathrm{probe}}$} = 0 case, both X-ray probes overlap in space and time, corresponding to the maximum contrast value that is analogous to the diagonal \emph{t$_{1}$~=~t$_{2}$} line in two-time XPCS correlation plots. For the \emph{t$_{\mathrm{pump}}$}~=~0 case, the experiment effectively reduces to a simple pump-probe XSVS experiment \cite{Grubel2007} that follows how fast the system decorrelates, but fails to capture photoinduced relaxation pathways and possible oscillatory dynamics. To map nonequilibrium dynamic behavior that may not display a definitive energetic pathway requires two coherent probes to follow the same optical excitation. For example, simulated speckle contrast maps in Fig.~\ref{Fig1}b,c show linear and exponential relaxation timescales while Fig.~\ref{Fig1}d shows a system that undergoes a glassy to jammed transition following optical excitation.

\subsection*{SiO$_{2}$ and Fe$_{3}$O$_{4}$}
	
Since X-ray speckle techniques couple to the domain configuration of the illuminated sample volume, it is necessary to ensure no instabilities exist that may be mistaken for dynamics. To verify the stability of the SDO, we first measure a known static sample of \mbox{700-\textmu m-thick} silica (SiO$_{2}$) powder in the pump-double-probe setup with an 800-nm optical excitation. The penetration depth of the optical excitation is less than 0.1\% of the sample thickness that would have a negligible effect on the speckle pattern. Figure~\ref{Fig2}a shows the average speckle pattern of the powder diffraction where \mbox{$t_{\rm probe} =1$~ps} while \emph{t$_{\mathrm{pump}}$} varies from $0 - 10$~ps. Taking a small region of the speckle pattern, marked by the white dotted box in Fig.~\ref{Fig2}a, we bin and average the speckle pattern by various \emph{t$_{\mathrm{pump}}$} delay ranges (see Fig.~\ref{Fig2}b-d). The sharp contrast of the speckle pattern independent of the number of shots, also depicted as a line cut in  Fig.~\ref{Fig2}e, is indicative of a static system. This confirms the stability of the SDO as any pointing instabilities from the experimental setup would manifest as a decay of the speckle contrast with increasing \emph{t$_{\mathrm{pump}}$}. In other words, we have ensured that the X-ray spot maintains a stable position on the sample and the overlap between the two X-ray pulses is constant.
	
After confirming the stability of the SDO, the same analysis was performed on our material of interest, Fe$_{3}$O$_{4}$, where we couple to the charge dynamics of the insulating trimeron phase by accessing the (001) diffraction peak at the Fe \emph{K}-edge resonant photon energy of~7.12~keV \cite{Lorenzo2008}. Figure~\ref{Fig2}f shows the average speckle pattern of the (001) charge diffraction peak with the same \emph{t$_{\mathrm{probe}}$} and \emph{t$_{\mathrm{pump}}$} conditions as for SiO$_{2}$. The speckle pattern binned by \emph{t$_{\mathrm{pump}}$} delays (see \mbox{Fig.~\ref{Fig2}g-i}) reveals a decay in speckle visibility as more shots are averaged, providing evidence that the optical excitation is altering the configuration of the trimeron network. The integrated speckle pattern binned by \emph{t$_{\mathrm{pump}}$} delays is analogous to `waterfall' plots in XPCS that provide qualitative evidence of dynamics in the system. 
	
Quantifying the dynamic timescale requires the use of spatial autocorrelation \cite{Sun_Thesis, Sun2020} to recover any lost contrast from the angular mismatch between the two probes. Figures~\ref{Fig3}a-c show the spatial autocorrelation signals for three measurements of Fe$_{3}$O$_{4}$ where \emph{t$_{\mathrm{pump}}$} varies from~$0 - 10$~ps while \emph{t$_{\mathrm{probe}}$} was fixed at 0.2, 1.0, and 5.0~ps. The angular mismatch offsets the two speckle patterns on the detector that is recovered by the side lobe of the autocorrelation. Therefore, the degree of correlation between the speckle patterns from the first and second X-ray pulses on the detector directly scales with the intensity of the side lobe, indicated by the white dotted box in Fig.~\ref{Fig3}a. Its intensity is given by 
 \begin{equation}
	\label{Eq2}
	A({t_\mathrm{probe}}) = 1 + (r-r^{2})\mu\beta_{0}(v|S(Q,t)|^{2} + (1-v)),
\end{equation}
where $r = I_{1}/(I_{1} + I_{2})$ is the ratio of a single pulse intensity to the summed intensity while the degree of overlap~$\mu$ between the two X-ray pulses follows $0 \leq \mu \leq 1$. An additional term $1-v$, corresponding to the fraction of static or pinned domains that prevent the system from fully decorrelating, is also included.
 
The presence of a finite side lobe intensity in \mbox{Fig.~\ref{Fig3}a-c} indicates the Fe$_{3}$O$_{4}$ system has not fully decorrelated within those temporal regimes, suggesting that a fraction of the domains are still static. Using a binned speckle pattern analysis (see Supplementary Information), we calculate a minimum fraction of domains that are static to be $1-v = 0.129$, constraining the normalized contrast to follow $\beta({t_{\mathrm{probe}} \rightarrow \infty})/\beta({t_{\mathrm{probe}} = 0}) = 0.129$. Vertical projections through the central spatial autocorrelation signals are shown in Fig.~\ref{Fig3}d, in line with simulations from~\cite{Sun_Thesis, Sun2020} that demonstrate the viability of this analytical technique for two-pulse XSVS. Taking into account the static domain fraction ($1-v$) and correcting for the $r$ and $\mu$ parameters (see Supplementary Information), we can extract the normalized time-resolved contrast plotted in Fig.~\ref{Fig3}e that is given by
    \begin{equation}
    \label{Eq3}
        \beta({t_\mathrm{probe}}) = \frac{A({t_\mathrm{probe}}) - 1}{(r-r^{2})\mu}.
	\end{equation}

 A loosely bound model based on a stretched exponential decay of the speckle contrast points to an exponent less than unity ($\alpha = 0.63$, see Fig. \ref{Fig3}e), that suggests the evolution of trimeron domains within 10~ps of the optical excitation exhibits liquid-like fluctuations, in contrast to jammed dynamics seen in equilibrium of many quantum materials \cite{Shpyrko2007, Zhang2017, Campi2022, Kukreja2018}. Resonant XPCS coupled to the Fe$_{3}$O$_{4}$ charge domains in equilibrium conditions show a compressed exponent of $\alpha \approx$ 1.5 \cite{Kukreja2018}, which points to the existence of a critical \emph{t$_{\mathrm{pump}}$} where dynamics evolve from glassy to jammed, as simulated in Fig.~1d. Additional \emph{t$_{\mathrm{probe}}$} measurement points will be needed to further constrain fitting parameters and to rule out oscillatory dynamics of the signal that may arise due to defects. 

\begin{figure}
	\includegraphics[width=\columnwidth]{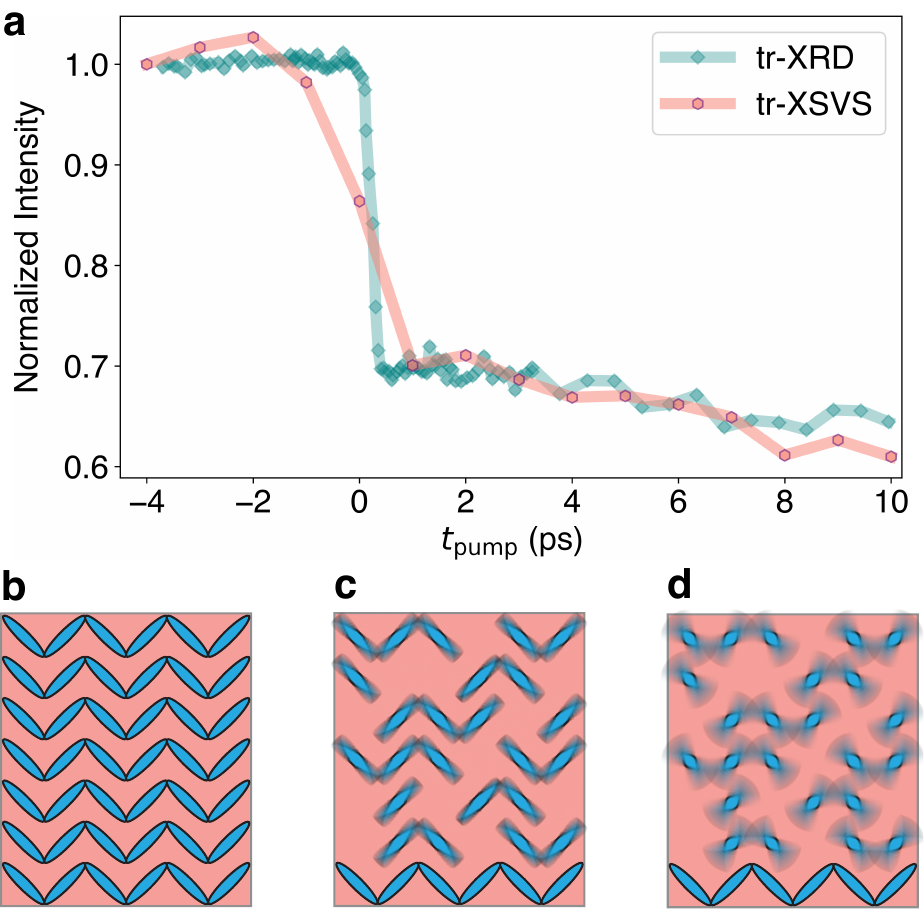}
	\caption{\label{Fig4} {\bf Ultrafast quench and fluctuation of the trimeron order parameter.} {\bf a} Time-resolved intensity of the  Fe$_{3}$O$_{4}$  (001) charge order peak from complementary \mbox{tr-XRD} and tr-XSVS experiments. The time delay refers to the pump-probe delay in tr-XRD and \emph{t$_{\mathrm{pump}}$} in tr-XSVS with \emph{t$_{\mathrm{probe}}$} = 0.2~ps. Schematics below show the trimeron network before optical excitation \emph{t$_{\mathrm{pump}} < 0$} (\textbf{b}), after optical excitation \emph{t$_{\mathrm{pump}} > 0$} and \emph{t$_{\mathrm{probe}}$} = 0.2~ps (\textbf{c}), and \emph{t$_{\mathrm{pump}} > 0$} and $t_{\rm{probe}} > 1.0$~ps (\textbf{d}).}
\end{figure}

Finally, we also conducted a pump-probe time-resolved X-ray diffraction (tr-XRD) measurement with the same experimental conditions but using `incoherent' X-rays to demonstrate the complementary nature of tr-XSVS and tr-XRD. The time-resolved integrated intensity of the resonant (001) diffraction peak is shown in Fig.~\ref{Fig4}a where the partial melting of the insulating domains after optical excitation can be seen in both measurements. However, due to the finite $t_{\rm{probe}} = 0.2$~ps in tr-XSVS, temporal resolution is lost within a $\Delta$\emph{t$_{\mathrm{probe}}$} window where the optical pump comes between the two X-ray probes, resulting in the summed intensity of a $t_{\mathrm{pump}} < 0$ and a $t_{\mathrm{pump}} > 0$ condition. Furthermore, the much lower photon flux of tr-XSVS due to the SDO requires coarser time bins to obtain sufficient signal-to-noise, as seen in Fig.~\ref{Fig4}a where the temporal resolution is limited to $t = 1$~ps. 
 
Figure~\ref{Fig4}a shows that the tr-XRD intensity undergoes an ultrafast quench of $\approx30\%$ within 1~ps followed by a slight thermalization decay within 10 ps. tr-XSVS provides nanoscale information, extracted from the spatial autocorrelation, that the surviving insulating domains are not static but fluctuating at these picosecond timescales. Taken altogether, we can formulate a real space picture that goes beyond a previous pump-probe experiment \cite{deJong2013}: As sketched in Fig \ref{Fig4}c, part of the trimeron network is melted after optical excitation, but the \emph{t$_{\mathrm{probe}}$} = 0.2 ps window is too short for the fluctuations of the order parameter to decorrelate the speckle pattern significantly. Once \emph{t$_{\mathrm{probe}} > 1$} ps, fluctuations of the trimeron network result in a considerable real-space reconfiguration that manifests as a larger decay of the contrast or speckle visibility (see Fig.~\ref{Fig4}d). 
	 
\section*{Discussion}
	
The structural and electronic orders behind the elusive Verwey transition of Fe$_{3}$O$_{4}$ have been investigated under equilibrium and nonequilibrium conditions. Equilibrium fluctuations of the charged-order domains have been studied with resonant XPCS~\cite{Kukreja2018, Hua2023}, while ultrafast phase separation of metallic and insulating domains was revealed in an optical pump -- \mbox{X-ray} probe experiment \cite{deJong2013}. Recently, an optical pump -- terahertz probe experiment revealed coherent soft modes of the trimeron order on picosecond timescales that are linked to charge dynamics along trimeron chains \cite{Baldini2020}. Our pump-double-probe tr-XSVS experiment demonstrates the unique ability to directly couple to these fluctuations in the time domain, revealing picosecond timescale dynamics. Since it is a time-domain technique that accesses \emph{S(Q,t)} instead of \emph{S(Q,$\omega$)}, tr-XSVS is ideal for capturing nonequilibrium dynamics induced by short optical pumps or pulsed magnetic fields that is hidden from frequency domain techniques. For example, using \mbox{tr-XRD} and time-resolved resonant inelastic X-ray scattering, evidence of in-plane diffusive dynamics of charge domains in La$_{2-x}$Ba$_{x}$CuO$_{4}$ was detected~\cite{Mitrano2019}. However, the exact relaxation timescales and potentially time-varying degree of diffusivity are not accessible using these techniques, but can in principle be extracted from \mbox{tr-XSVS}. 
	
We also demonstrate how tr-XSVS complements \mbox{tr-XRD} experiments by providing the details of local fluctuations of the order parameter that can have far reaching implications for emergent behaviors in quantum materials. For example, hidden phases induced via nonthermal pathways have been discovered in materials such as VO$_{2}$ \cite{Cavalleri2001}, V$_{2}$O$_{3}$ \cite{Singer2018} and 1\emph{T}-TaS$_{2}$ \cite{Stojchevska2014, Domrose2023} that have transient lifetimes on the picosecond timescales. \mbox{tr-XRD} experiments have confirmed these hidden states, but can not reveal dynamical information beyond changes in the average structure \cite{Hua1TTaS2}. With the pump-double-probe experimental scheme introduced here, the `pump' grants access to the hidden state while the two subsequent probes with variable delay times couple to the dynamics that can uncover clues into the origins of the metastability. 

We benchmark the ultrafast fluctuation timescale of the trimeron network of Fe$_{3}$O$_{4}$ where the dynamics within 10~ps of the optical excitation are similar to a supercooled liquid phase \cite{Ruta2012, Tong2018}. In other words, the optical quench depins the electronic system in a way that the charge domain dynamics become more particle-like with a broad distribution of timescales that stretches the exponent. This deviates from the Arrhenius trend of jammed dynamics seen under equilibrium conditions~\cite{Kukreja2018}. Our findings suggest that the fundamentally different dynamics of an electronic order parameter may play a vital role in driving emergent behaviors under nonthermal external stimuli. Resolving the full relaxation pathway will require higher resolution in terms of \emph{t$_{\mathrm{pump}}$}, where the current statistics is still insufficient to achieve the required signal-to-noise ratio. Split-and-delay setups employing a transmission grating that produces colinear \mbox{X-ray} pulses will obviate the need for analytical methods of correcting for an angular mismatch between the two pulses \cite{Li2021}. This will yield an increase of the maximum initial contrast by a factor of two where the dynamics from a decay in contrast can be extracted directly from photon counting  \cite{Hruszkewycz2012, Seaberg2017}. Furthermore, improvements to the experimental setup and increased coherent flux from MHz X-ray FELs will enable sufficient statistics to construct speckle-contrast maps, the key to unveiling ultrafast spatio-temporal dynamics of order parameters under nonequilibrium conditions.
		
\section*{Methods}
	
\noindent\textbf{Sample Fabrication.} A 320-nm Fe$_{3}$O$_{4}$ thin film was deposited on a (001)-oriented MgO substrate using reactive magnetron sputtering in an Ar/O$_{2}$ environment at 500$^{\circ}$ C. The total pressure of the sputtering chamber was 2.4~mTorr while the partial pressure of O$_{2}$ was~0.1~mTorr. Prior to deposition, the MgO substrate was annealed at 500~$^{\circ}$C for one hour under vacuum in order to improve the film quality. The sample was used for both tr-XRD and tr-XSVS.\\
		
\noindent\textbf{tr-XSVS and tr-XRD.} The pump-double-probe XSVS experiment was carried out at the X-ray Correlation Spectroscopy~(XCS) endstation of LCLS~\cite{AlonsoMori2015} using the SDO setup~\cite{Zhu2017}. An \mbox{800-nm} laser at a nominal fluence of~1.5~mJ/cm$^{2}$ and a full width at half maximum (FWHM) of 35~fs was used to excite the sample. Two subsequent coherent X-ray pulses with a FWHM of 50~fs tuned to the Fe \emph{K}-edge resonant energy (7.12~keV) were used to probe the (001) charge order peak of Fe$_{3}$O$_{4}$. The X-ray spot size was approximately \mbox{10 x 10 \textmu m}$^{2}$. The penetration depth of the \mbox{800-nm} laser is $\approx 118$~nm~\cite{deJong2013} while that of the X-rays is $\approx440$~nm. A liquid nitrogen cryojet was used to cool the sample below the Verwey transition to $T\approx 95$~K where charge order emerges. The Fe$_{3}$O$_{4}$ (001) speckle pattern was captured on one of four ePix100 detector panels~\cite{Carini2016} located 8~m from the sample while the SiO$_{2}$ powder diffraction covered all four detector panels. The complementary tr-XRD (pump-probe) experiment was performed at the Bernina endstation of SwissFEL~\cite{Ingold2019} using a monochromator at the same X-ray energy and approximately the same pulse duration, but with a spot size of \mbox{50 x 50 \textmu m}$^{2}$. An equivalent 800-nm pump laser setup at an incident fluence of~1.5~mJ/cm$^{2}$ was used to excite the sample that was also maintained at $T\approx 95$~K by a liquid nitrogen cryojet. A time tool correction with 50 fs bin sizes was used that defines an overall time resolution of $\approx80$~fs. A three-module JungFrau~1.5M detector~\cite{Mozzanica2018} located $\approx1$~m from the sample was used to detect the Fe$_{3}$O$_{4}$ (001) peak. In both experiments, the \mbox{800-nm} laser was nearly colinear with the X-rays in a horizontal scattering geometry, but the XCS~experiment ran at a repetition rate of 120~Hz with vertical polarization while the Bernina~experiment was carried out at 100~Hz with horizontal polarization. 

\section*{Acknowledgements}

Use of the Linac Coherent Light Source (LCLS), SLAC National Accelerator Laboratory, is supported by the U.S. Department of Energy, Office of Science, Office of Basic Energy Sciences under Contract~\mbox{DE-AC02-76SF00515}. We acknowledge the Paul Scherrer Institute, Villigen, Switzerland for provision of free-electron laser beamtime at the Bernina instrument of the SwissFEL Aramis branch and synchrotron radiation beamtime at the X04SA beamline of the Swiss Light Source. We would also like to acknowledge technical assistance by A.~R.~Oggenfuss during the preparation of the SwissFEL Bernina experiment. This research has received funding from the European Union’s Horizon 2020 research and innovation programme under the Marie Sklodowska-Curie Grant Agreement 884104 \mbox{(PSI-FELLOW-III-3i)}, and in part by the Swiss National Science Foundation (SNSF) Grant 213148. M.C. and A.G.M. acknowledge funding from the European Research Council under the European Union’s Horizon~2020 research and innovation program, within Grant Agreement~810451 (HERO). P.R., N.Z.H., M.M., S.T.B., S.J. and R.K. acknowledge support from U.S. Department of Energy, Office of Science, Basic Energy Sciences under Award Number DE-SC0022287. B.K.S., E.S.L., and O.G.S were supported as part of the Quantum Materials for Energy Efficient Neuromorphic Computing (Q-MEEN-C) Energy Frontier Research  Center (EFRC), funded by the U.S. Department of Energy, Office of Science, Basic  Energy Sciences under Award~\mbox{DE-SC0019273}.


\bibliography{Main/apssamp}
\clearpage


\preprint{APS/123-QED}
	
	\title{Supplementary Information: \\\textit{Imaging ultrafast electronic domain fluctuations with X-ray speckle visibility}}
\author{N. Hua}
	\email{nelson.hua@psi.ch}
	\affiliation{PSI Center for Photon Science, Paul Scherrer Institute, 5232 Villigen PSI, Switzerland}%
	\author{Y. Sun}
	\affiliation{Linac Coherent Light Source, SLAC National Accelerator Laboratory, Menlo Park, California 94025, USA}
	\author{P. Rao}
	\author{N. Zhou Hagstr\"{o}m}
	\affiliation{Department of Materials Science Engineering, University of California Davis, Davis, California 95616, USA}
	\author{B. K. Stoychev}
	\author{E. S. Lamb}
	\affiliation{Department of Physics, University of California San Diego, La Jolla, California 92093, USA}
	\author{M.~Madhavi}
	\author{S.~T. Botu}
	\affiliation{Department of Materials Science Engineering, University of California Davis, Davis, California 95616, USA}
	\author{S.~Jeppson}
	\affiliation{Department of Materials Science Engineering, University of California Davis, Davis, California 95616, USA}
	\author{M. Cl{\'e}mence}
	\author{A. G. McConnell}
	\affiliation{PSI Center for Photon Science, Paul Scherrer Institute, 5232 Villigen PSI, Switzerland}%
	\affiliation{Laboratory for Solid State Physics and Quantum Center, ETH Zürich, 8093 Zürich, Switzerland}
	\author{S.-W. Huang}
	\author{S.~Zerdane}
	\author{R. Mankowsky}
	\author{H. T. Lemke}
	\author{M. Sander}
	\affiliation{PSI Center for Photon Science, Paul Scherrer Institute, 5232 Villigen PSI, Switzerland}
	\author{V. Esposito}
	\author{P. Kramer}
	\author{D. Zhu}
	\author{T.~Sato}
	\author{S. Song}
	\affiliation{Linac Coherent Light Source, SLAC National Accelerator Laboratory, Menlo Park, California 94025, USA}
	\author{E. E. Fullerton}
	\affiliation{Center for Memory and Recording Research, University of California San Diego, La Jolla, California 92093, USA}
	\author{O. G. Shpyrko}
	\affiliation{Department of Physics, University of California San Diego, La Jolla, California 92093, USA}
	\author{R. Kukreja}
	\email{rkukreja@ucdavis.edu}
	\affiliation{Department of Materials Science Engineering, University of California Davis, Davis, California 95616, USA}
	\author{S. Gerber}
	\email{simon.gerber@psi.ch}
	\affiliation{PSI Center for Photon Science, Paul Scherrer Institute, 5232 Villigen PSI, Switzerland}%
	\date{\today}
\maketitle

\section{XSVS data reduction}

\begin{figure}[t]
		\includegraphics[width=\columnwidth]{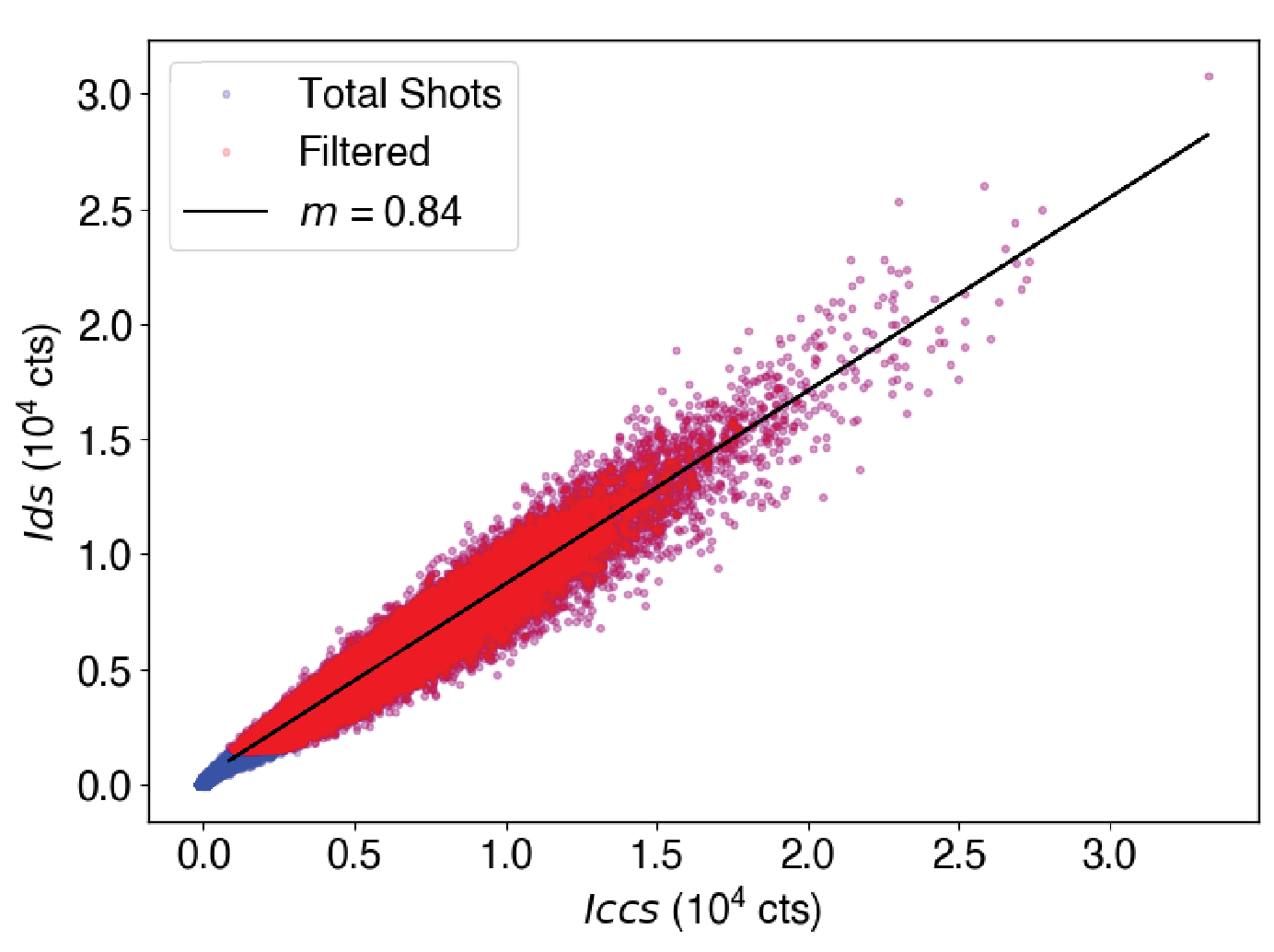}
		\caption{\label{SF1} The \textit{Ids} and \textit{Iccs} monitors follow the shot-to-shot intensity of the  X-ray pulses from the two SDO branches. For each measurement, linear correlation (with slope $m$) is confirmed to ensure similar intensities amongst the two lines. Low intensity counts are filtered out from the final dataset.}
	\end{figure}

\begin{figure}[t]
		\includegraphics[width=\columnwidth]{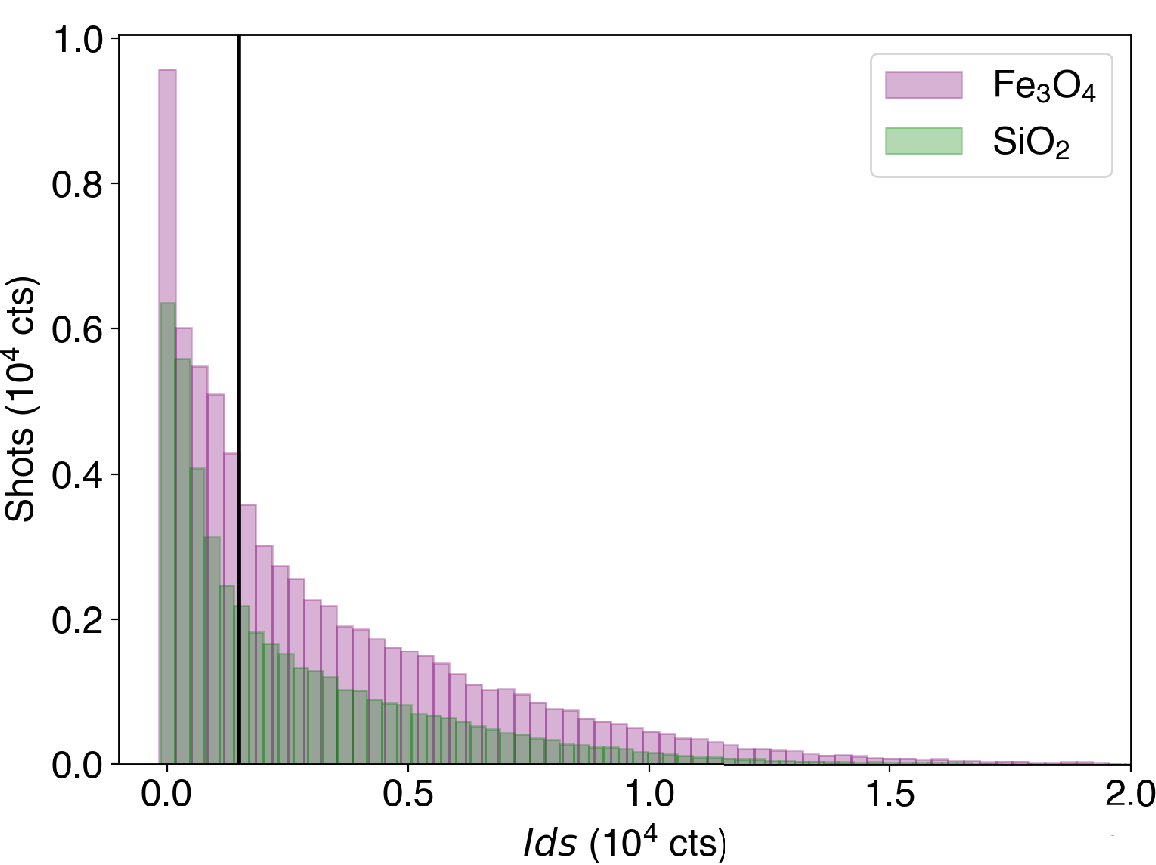}
		\caption{\label{SF2} Single-shot intensity histogram of a typical measurements of the SiO$_{2}$ powder ($\approx$ 40,000 shots) and the Fe$_{3}$O$_{4}$ sample ($\approx$ 70,000 shots). A threshold of 1500 cts (black line) was used to filter out low intensity incident pulses in order to optimize the signal-to-noise of the speckle pattern.}
	\end{figure}

In a double-probe XSVS setup, the intensities between the two X-ray pulses after the split-and-delay optics (SDO) branch must be approximately equal to ensure mutual coherence~\cite{Li2021}. Therefore, correlations between the main and delayed branches of the SDO setup are verified with intensity monitors \textit{Iccs} and \textit{Ids}, respectively (see Fig.~\ref{SF1}). A typical histogram of the intensities is shown in Fig.~\ref{SF2} for both a SiO$_{2}$ and Fe$_{3}$O$_{4}$ scan where a threshold of 1500 counts was used to filter out low incident X-ray intensities. A detector mask was also applied to filter out dead or hot pixels on the ePix100 detector panels that would affect the subsequent spatial autocorrelation analysis. Finally, a greedy-guess based droplet algorithm was applied to localize and discretize the counts into integer photon hits on the detector~\cite{sun2020accurate}. 

\section{Split-and-delay alignment and stability}

\begin{figure}[!ht]
		\includegraphics[width=\columnwidth]{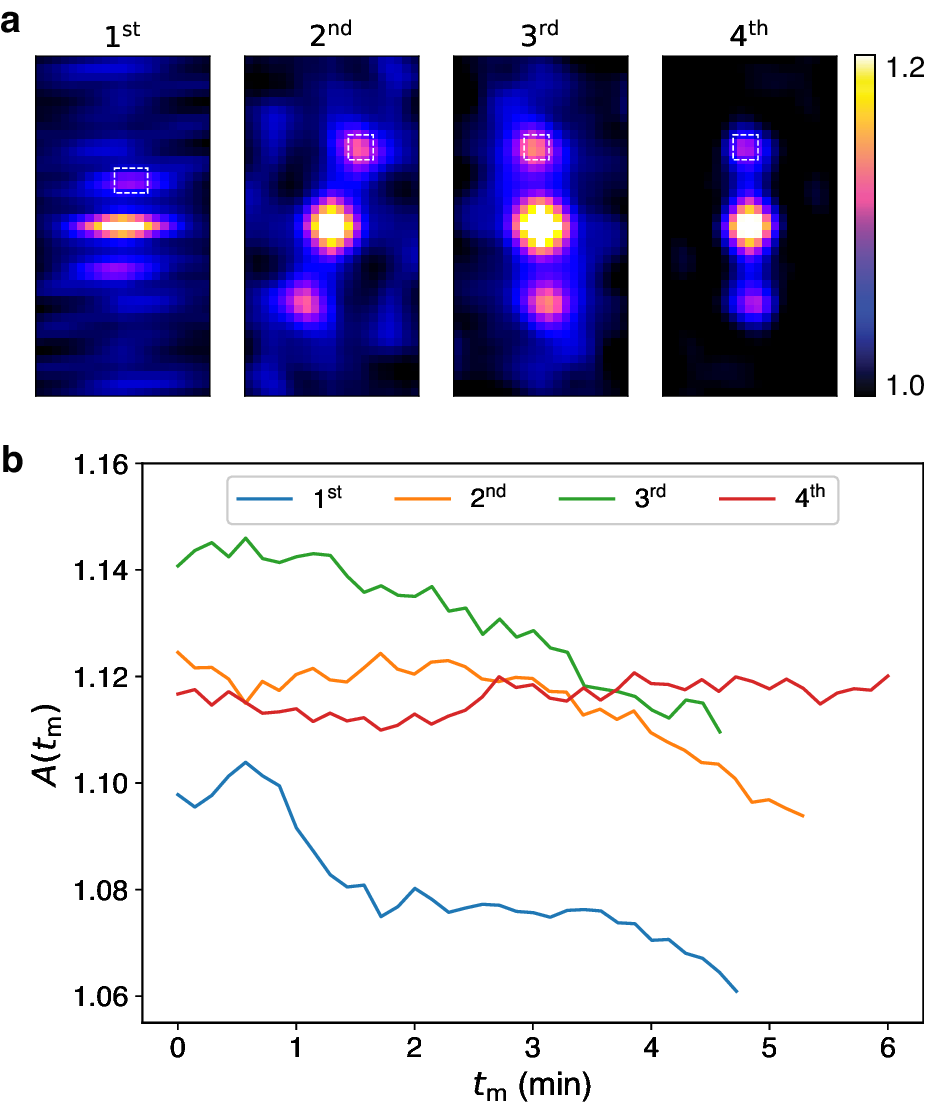}
		\caption{\label{SF3} {\bf {a}} The spatial autocorrelation side lobes, coming from the static SiO$_{2}$ powder speckle pattern during alignment, are marked by the white dotted boxes. Between the first and second measurement, the (horizontal, vertical) slits were adjusted from (0.3, 0.1) to (0.1, 0.3)~mm. From the second to the fourth measurements, the alignment of the angular mismatch between the two X-ray pulses was adjusted to be orthogonal to the scattering plane of the subsequent Fe$_{3}$O$_{4}$ sample. {\bf {b}} The shot-to-shot stability can be verified by following the autocorrelation lobe intensity $A(\it{t}_{\mathrm{m}})$, defined by the integrated region of the white dotted boxes in \textbf{a}, as a function of \emph{t$_{\mathrm{m}}$}. By the fourth measurement, a stable correlation value that reflects the static SiO$_{2}$ sample is reached.}
	\end{figure}

An ideal speckle visibility experiment relying on multiple X-ray pulses consists of completely identical coherent pulses propagating colinearly with full spatial overlap. Ensuring the stability of the SDO setup is therefore crucial for experiments that are sensitive to the exact domain configuration in the system. As already shown in Fig.~2 of the main text, the SDO stability is qualitatively confirmed by the speckle visibility of the static SiO$_{2}$ powder sample. The spatial autocorrelation analysis can also be used to align and quantify the shot-to-shot stability of the spatial overlap between the two X-ray probes. Figure~\ref{SF3}a shows the spatial autocorrelation signal of the SiO$_{2}$ powder speckle pattern for four measurements. The side lobe intensity profiles are used as a guide to adjust the vertical and horizontal slit sizes and to align the angular mismatch to be orthogonal to the scattering plane of the subsequent Fe$_{3}$O$_{4}$ experiment. The stability of the setup can be checked by following the intensity of the side lobe, $A(\it{t}_{\mathrm{m}})$, as a function of the measurement time $\it{t}_{\mathrm{m}}$ (see Fig.~\ref{SF3}b). The constant side lobe intensity of the fourth measurement indicates that not only is the sample static, but the spatial overlap between the two X-ray pulses is also stable.  

\section{measurement scheme}

\begin{figure}[!ht]
		\includegraphics[width=\columnwidth]{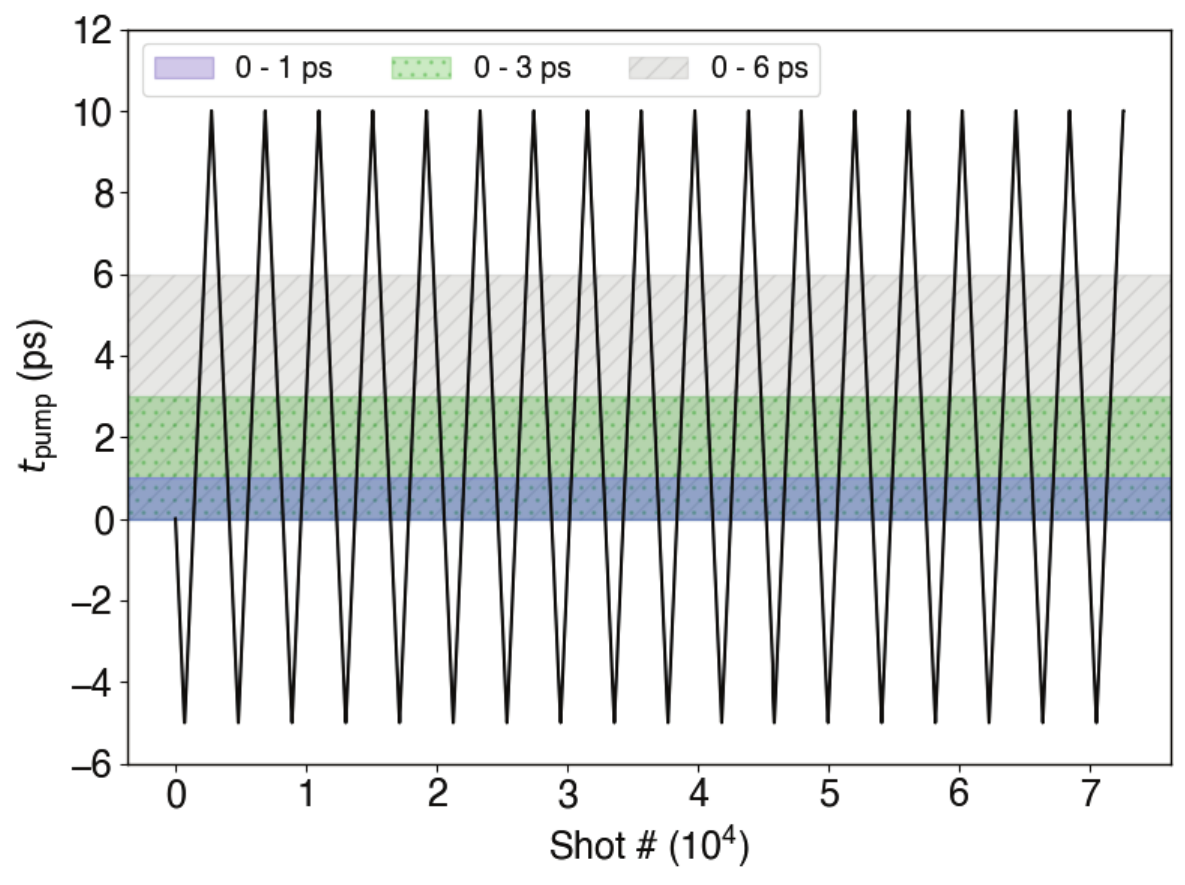} 
		\caption{\label{SF4} `Flying' delay scan where \emph{t$_{\mathrm{pump}}$} varies between \mbox{$-5$ to 10~ps} while $t_{\mathrm{probe}} = 1$~ps. The highlighted regions correspond to the temporal windows of the binned speckle patterns in Fig.~2 of the main text.}
	\end{figure}

\begin{figure}
		\includegraphics[width=.9\columnwidth]{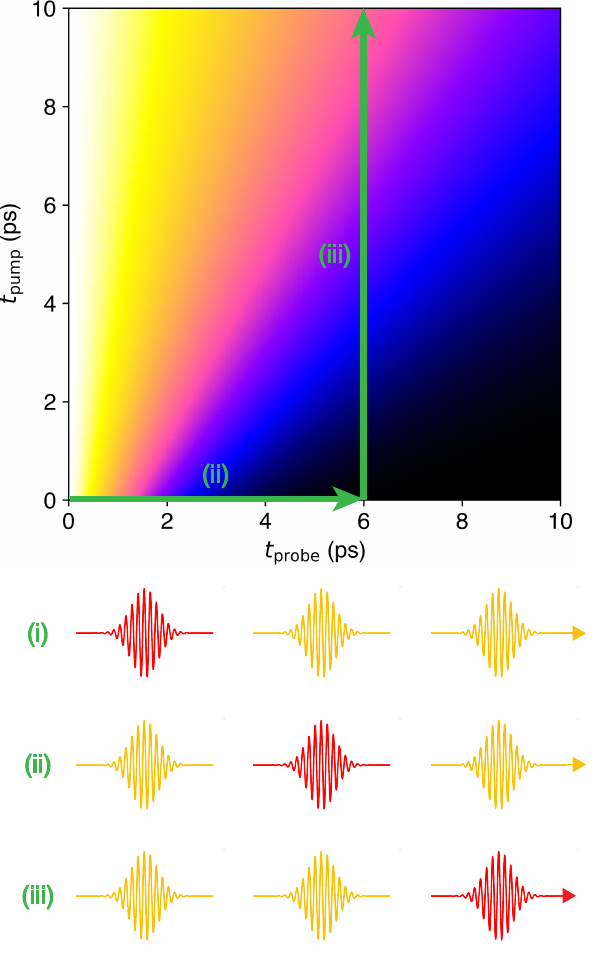}
		\caption{\label{SF5} The experimental measurement scheme can be divided into three segments and proceeds as follows: (i) both X-ray pulses hit the sample before the optical laser excitation (\emph{t$_{\mathrm{pump}}$} $< 0$ and $|$\emph{t$_{\mathrm{pump}}$} $| >$ \emph{t$_{\mathrm{probe}}$}), (ii) the optical pump comes in between the two X-ray probes that reduces to a probe-pump-probe experiment (\emph{t$_{\mathrm{pump}}$}  $< 0$ and $|$\emph{t$_{\mathrm{pump}}$} $| <$ \emph{t$_{\mathrm{probe}}$}), and (iii) the optical pump excites the sample followed by the two X-ray probes at subsequent time delays (\emph{t$_{\mathrm{pump}}$} $> 0$). }
	\end{figure}

Since the stability between the two coherent X-ray pulses is critical in speckle visibility experiments, the delay stage between the two probes is kept constant during a measurement while the delay between the optical pump and the first X-ray pulse is varied as a `flying' scan (see~Fig.~\ref{SF4}). The scanning mechanism can thus be separated into three segments (see Fig.~\ref{SF5}), where an intermediate segment reduces to effectively a probe-pump-probe experiment. Maintaining the \emph{t$_{\mathrm{pump}}$} range for each scan while adjusting the \emph{t$_{\mathrm{probe}}$} delays between measurements allows us to reconstruct the full speckle contrast maps in the pump-double-probe XSVS experiment.
 
\section{Contrast considerations and extracting model parameters}

\begin{figure}[!ht]
    \includegraphics[width=\columnwidth]{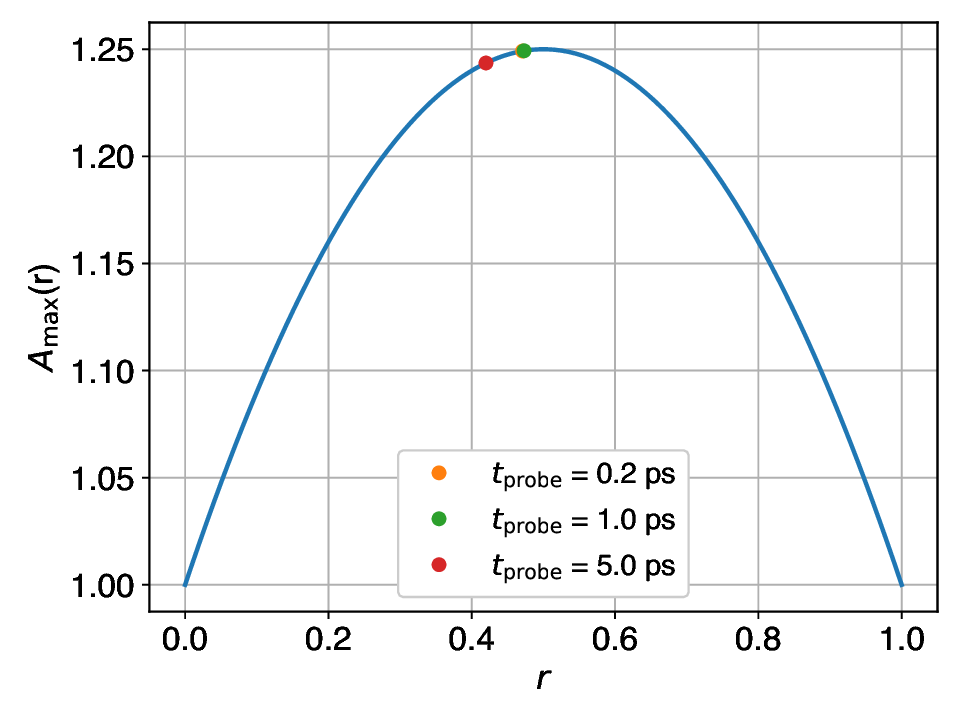}
    \caption{\label{SF6} Maximum side lobe intensity $A_{\rm max}$ as a function of~$r$ showing how the initial contrast is reduced due to an imperfect intensity ratio between the two X-ray probes. Overlaid is the intensity reduction of each of the experimental \emph{t$_{\mathrm{probe}}$}~delays (see Table I).}
\end{figure}

The autocorrelation side lobe intensity $A(t_{\mathrm{probe}})$ is proportional to the intermediate scattering function $S(Q,t)$ 
\begin{equation}
    \label{Eq1}\tag{S1}
	A({t_\mathrm{probe}}) = 1 + (r-r^{2})\mu\beta_{0}(v|S(Q,t)|^{2} + (1-v)),
\end{equation}
where $r$ and $\mu$ are experimental parameters \cite{Sun_Thesis}.
~\mbox{$r=\frac{I_{1}}{I_{1} + I_{2}}$} is the ratio of the single-pulse to the double-pulse intensity. $0 \leq \mu \leq 1$ is the overlap fraction between the two X-ray pulses. In the ideal case where $r = 0.5$ and $\mu = 1$, the maximum side lobe intensity is 1.25, corresponding to a maximum contrast difference of 0.25. $r$ can be obtained from the \textit{Ids} and \textit{Iccs} monitors (see Fig.~\ref{SF1}). The slope~$m$ of the correlation corresponds to the ratio of the intensities that can be equivalently derived by $r_{\rm b} = \frac{1}{1+m}$. The subscript b refers to the intensities `before the sample'. A correction from an additional intensity variation between the two X-ray probes originating from the sample must also be made. The tr-XRD signal shown in Fig.~4a of the main text reveals a slight decay of the diffraction intensity following \emph{t$_{\mathrm{pump}} >$} 0 ps, supposedly due to heating effects. Thus, larger \emph{t$_{\mathrm{probe}}$} delays will also see a larger intensity difference between the two \mbox{X-ray} probes. A linear fit was applied to the tr-XRD signal after \emph{t$_{\mathrm{pump}} >$}~0~ps from which the ratio of intensities was extracted, designated as $r_{\rm a}$ where the subscript a refers to `after the sample'. The effective $r$~parameter due to multiple intensity variation effects can be described by 
\begin{equation}
    \label{Eq3}\tag{S2}
    r_{eff} = 2^{N-1}\prod_{i=1}^{N}r_{i},
\end{equation}
which can be calculated from  $r_{\rm b}$ and $r_{\rm a}$ (see Table I). Figure~\ref{SF6} shows the decrease in the optimal contrast due to these intensity variations. 

\begin{table}[t]
  \centering
  \begin{tabular}{|c|c|c|c|}
    \hline
    {  $t_{\mathrm{probe}}$ (ps)} & { $r_{\rm b}$} & { $r_{\rm a}$} & { $r_{\rm eff}$} \\
    \hline
    0.2 & 0.471 & 0.499& 0.470\\
    \hline
    1.0  & 0.473& 0.498 & 0.471\\
    \hline
    5.0 & 0.420 & 0.489 & 0.411\\
    \hline
  \end{tabular}
  \caption{$r$ parameters at different X-ray pulse time delays $t_{\rm probe}$, extracted from the experiment.}
  \label{tab:my_table}
\end{table}

\begin{figure}[!ht]
    \includegraphics[width=\columnwidth]{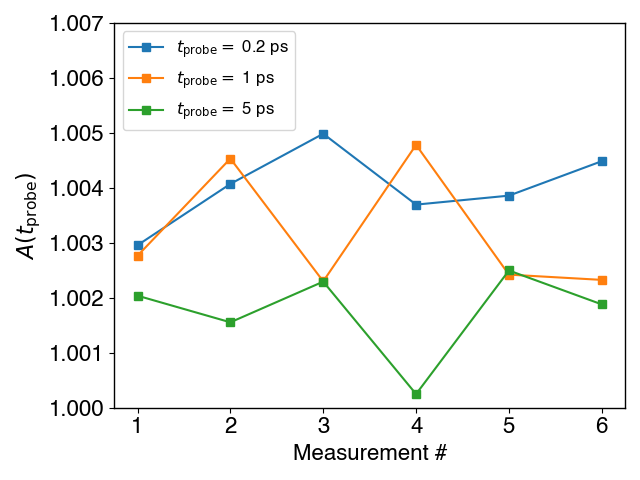}
    \caption{\label{SF7} Side lobe intensity $A(t_{\rm probe})$ of six successive Fe$_{3}$O$_{4}$ measurements. The spatial overlap was re-aligned before each first measurement.}
\end{figure}

$\mu$ is not possible to monitor on a shot-to-shot basis unless a static speckle pattern can be simultaneously captured. However, our experimental procedure and various observables allow us to approximate its value. The steady speckle contrast from the static SiO$_{2}$ powder sample in Fig.~2 of the main text qualitatively shows that the spatial overlap between the two X-ray probes remains constant for the duration of a scan. For the Fe$_{3}$O$_{4}$ sample, the spatial overlap was re-aligned followed by six \mbox{$\approx10$-minute} measurements for each of the \emph{t$_{\mathrm{probe}}$} delay times. The side lobe height of each measurement is shown in Fig.~\ref{SF7}. If the spatial overlap drifted significantly over the measurement time, a decay of the side lobe intensity due to decreasing $\mu$ would have been seen. Therefore, the value of $\mu$ is unchanging. To approximate the value of $\mu$, we can use the static SiO$_{2}$ powder speckle measurement in Fig.~2 where $t_{\rm{probe}} = 1$~ps, corresponding to~$r_{\rm eff} = 0.471$. We can derive the static contrast by 

\begin{equation}
    \label{Eq4}\tag{S3}
    \beta_{0} = \frac{\sigma^{2}(M)}{\bar{M}^{2}[r^{2}+(1-r)^{2}]},
\end{equation}
where $\sigma^{2}(M)$ is the variance of the speckle pattern~$M$. Inputting $\beta_{0}$ and the average side lobe height~\mbox{$A(t_{\rm m}) = 1.116$} from Fig. \ref{SF3} into Eq.~(3) of the main text, we extract $\mu$ = 0.69.

Finally, an imperfect penetration depth overlap between the optical pump and X-ray probe (see Methods) exists where we expect the bottom portion of the film to see the two X-ray pulses, but not the effects of the optical pump. To estimate the fraction of the film that is unperturbed, we randomly averaged all speckle patterns within a measurement into $N$ = 10 bins, each representing a possible mode \emph{M}. The contrast is estimated by taking the normalized variance of $(\sum_{i}^{N}M_{i})/N$ where a fully decorrelated system follows $1/N$, in other words, $M$ = 10 different modes for $N$ = 10 bins. Here, the intensity of each speckle pattern can be decomposed into $M_{i} = I_{\rm s}I_{\rm sf}$ where $I_{\rm sf}$ is the Gaussian envelope of the Fe$_{3}$O$_{4}$ (001) charge peak from the structure factor, and $I_{\rm s}$ is the overlaid speckle contrast. $I_{\rm sf}$ was obtained from a Gaussian filter with a standard deviation of three pixels that was subsequently divided out to obtain the pure speckle component $I_{\rm s}$ that was used to calculate the normalized contrast. The same procedure was also applied before taking the spatial autocorrelation, shown in Fig.~3 of the main text.

\begin{figure}[b]
    \includegraphics[width=\columnwidth]{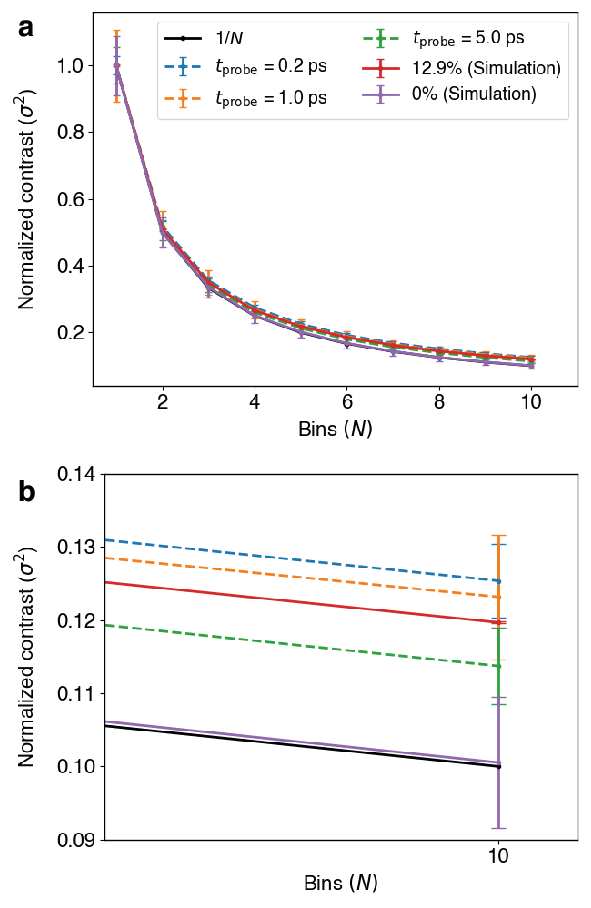}
    \caption{\label{SF8} {\bf a} The decay of the normalized contrast for speckle patterns that have been randomly averaged into \mbox{$N = 10$} bins is shown by dashed lines. Simulations of completely decorrelated speckle patterns (0$\%$) and one where a fraction (12.9$\%$) of the domains remains static to the external laser excitation are shown by solid lines. The black line depicts a $1/N$ behaviour. {\bf b} Zoomed in view at $N = 10$, revealing that a fraction of the charge domains of Fe$_{3}$O$_{4}$ is indeed static, likely due to the penetration depth mismatch of the 800-nm and X-ray pulses in the experiment.}
\end{figure}

Figure~\ref{SF8} shows our calculation for the various \emph{t$_{\mathrm{probe}}$} delays. Simulations of speckle patterns with contributions from static domains show that at least 12.9\% of the domains after \emph{t$_{\mathrm{pump}} >$} 0 are static. This value was obtained by taking the average normalized variance at \mbox{$N = 10$} for all \emph{t$_{\mathrm{probe}}$} delays and allows us to constrain the maximum and minimum contrast from \mbox{$\frac{\beta(t \rightarrow \infty)}{\beta(t_{\mathrm{pump}} < 0)} = 0.129$}. As seen in Fig. \ref{SF8}b, there is a general agreement across the \emph{t$_{\mathrm{probe}}$} delays, though the \emph{t$_{\mathrm{probe}}$}~=~0.2 and 5.0 ps data points are slightly outside the error bars. This suggests there could be a minimal effect due to the first X-ray probe, but this leads to a correction that would only further stretch the exponential fit shown in Fig.~3e of the main text. Simulations of completely decorrelated speckle patterns follow $1/N$ as expected. This analysis is important for extracting the contrast decay from the intermediate scattering function in XSVS to determine if any static or pinned domains exist that would change the baseline contrast value. 



\end{document}